# Achieving Effective Renormalization Scale and Scheme Independence *via* the Principle of Observable Effective Matching

Farrukh A. Chishtie[1,2]*

[1]*Department of Applied Mathematics, The University of Western Ontario, London, ON, Canada,* [2]*Spatial Informatics Group, LLC, Pleasanton, CA, United States*

In this work, we explicate a new approach for eliminating renormalization scale and scheme (RSS) dependence in observables. We develop this approach by matching RSS-dependent observables (such as cross-sections and decay rates) to a theory which is independent of both these forms of dependencies. We term the fundamental basis behind this approach as the principle of observable effective matching (POEM), which entails matching of a scale- and scheme-dependent observable with the fully physical scale (PS) and dynamical scale-dependent theory at loop orders at which RSS independence is guaranteed. This is aimed toward achieving so-called "effective" RSS-independent expressions as the resulting dynamical dependence is derived from a particular order in RSS-dependent perturbation theory. With this matching at a PS at which the coupling (and masses) is experimentally determined at this scale, we obtain an "effective theoretical observable (ETO)", a finite-order RSS-independent version of the RSS-dependent observable. We illustrate our approach with a study of the cross-section ratio $R_{e^+e^-}$ for $e^+e^- \to$ hadrons, which is demonstrated to achieve scale and scheme independence utilizing the three- and four-loop order $\overline{MS}$ scheme expression in QCD perturbation theory via matching at both one-loop and two-loop orders for obtaining the ETO. With two-loop matching, we obtain an ETO prediction of $\frac{3}{11}R^{eff}_{e^+e^-} = 1.052431^{+0.0006}_{-0.0006}$ at $Q = 31.6 GeV$, which is in excellent agreement with the experimental value of $\frac{3}{11}R^{exp}_{e^+e^-} = 1.0527^{+0.005}_{-0.005}$. Given its new conceptual basis, ease of use, and performance, we contend that POEM be explored in its application for obtaining ETOs for predicting RSS-independent observables across domains of high-energy theory and phenomenology as well as other areas of fundamental and applied physics, such as cosmology and statistical and condensed matter physics.

# 1 INTRODUCTION

In perturbative quantum field theory (QFT), singularities are encountered in theoretical expressions of physical observables which require renormalization techniques. Rendering finiteness to such quantities, such as physical cross-sections and decay rates, introduces renormalization scale and scheme (RSS) dependencies. These dependencies ultimately lead to theoretical uncertainties in predictions. This is especially problematic in the case of perturbative quantum chromodynamics (pQCD), wherein the strong coupling constant is large and RSS dependencies can lead to higher theoretical uncertainties. Hence, the main motivation of this work is the elimination of the RSS dependencies, which is aimed to lead to theoretical predictions with higher accuracy. This is demonstrated in the case of the QCD cross-section ratio $R_{e^+e^-}$. At a broader level, this work aims to bridge conventional renormalization approaches with effective field theoretical techniques to render a fully finite observable without any RSS ambiguities.

The issue of RSS ambiguities is treated via the principle of minimal sensitivity (PMS) [1–4]. Other approaches to reducing and/or eliminating scheme dependence include the approaches of effective charge [5, 6], Brodsky–Lepage–Mackenzie (BLM) [7–9], renormalization group (RG) summation [10], RG summation with RS invariants [11–14], complete renormalization group improvement (CORGI) [15–17], the principle of maximal conformality (PMC) [18–23], and sequential extended BLM (seBLM) [24]. In this work, we detail a new and alternate approach toward achieving RSS independence, inspired by effective field theory (EFT) techniques of matching. Our work is also surmised on the observation that at increasing order in the expansion of a small expansion parameter, along with convergence, such dependencies would be reduced and the penultimate expression would have full dependence only on dynamical and physical scales. As such, it is based on what we conceptualize as the "principle of observable effective matching" (POEM). Since this principle is applicable for achieving RSS independence only up to a fixed order in perturbation theory, we use the term "effective" for the observable having only physical scale (PS) and dynamical scale dependence, which is also the scale at which matching occurs. We term the derived perturbative expression as an "effective theoretical observable (ETO)", which implies these caveats. The study is outlined as follows: we introduce POEM next and then work out the RSS-independent quantum chromodynamics (QCD) cross-section ratio $R_{e^+e^-}$ via POEM at the one- and two-loop matching with an RSS-independent scheme in the following sections.

# 2 THE PRINCIPLE OF OBSERVABLE EFFECTIVE MATCHING

Physical observables computed to all orders in perturbative QFT within various RSS approaches are expected to result in scale and scheme independence if results at all orders are computed and hence must be the same in this limit. This is the starting point for POEM which can be expressed as follows:

$$\lim_{n \to \infty} O_n^{(i)}(a_n^{(i)}(\mu^{(i)}), m_n^{(i)}(\mu^{(i)}), \mu^{(i)}, Q, M) = \lim_{l \to \infty} O_l^{(j)}(a_l^{(j)}(\mu^{(j)}),$$
$$m_l^{(j)}(\mu^{(j)}), \mu^{(j)}, Q, M) = O_{full}(a(Q), m(M), Q, M). \quad (1)$$

Here, $O$ denotes perturbative contributions to a physical observable such as in a decay rate or a cross-section computed in two schemes denoted by superscripts $i$ and $j$ at some loop orders $n$ and $l$, respectively, while $a$ and $m$ denote the coupling and mass, respectively, tied to the respective renormalization scales $\mu^i$ and $\mu^j$. $Q$ and $M$ denote dynamical PS. **Eq. 1** can be further generalized by including multiple couplings and masses as well as factorization scales and momentum fractions. This equation nevertheless implies that if $O$ is computed at a certain large limit of perturbation theory, this results in an RSS-independent result with dependence only on $Q$ and $M$. While it is rather challenging to find expressions to all orders in perturbation theory, we consider such to be an RSS-independent theory or full theory whereby all implicit and explicit renormalization scales within any scheme cancel, and the equalization across in **Eq. 1** implies an overall RSS independence, leading to an independent observable termed as $O_{full}$ dependent only on a scale-independent coupling and mass, as well as physical scales, $Q$ and $M$.

While in **Eq. 1**, RSS independence is achieved for an observable computed at all orders in perturbation theory, for practical purposes, and as criteria to derive an effective version of such an all-order observable, RSS independence holds true trivially at the tree level (when no quantum corrections exist), and this is also valid at the some $r$-loop level at a physical scale (typically, this is at one- and two-loop levels). With these properties in mind, we propose the POEM condition which can be applied to achieve an effective RSS independence at an order-by-order basis for truncated expressions of a physical observable as follows:

$$O_n^{(k)}(a_n^{(k)}(Q^*), m_n^{(k)}(Q^*), Q^*, Q, M)$$
$$= O_{eff}^{(r)}(a_{eff}(Q^*), m_{eff}(Q^*), Q^*, Q, M). \quad (2)$$

Here, $O_{eff}$ is conceptualized in a finite-order perturbative physical representation termed as the ETO which is truncated at the $r$-loop order in order to be consistent with RSS independence and is dependent only on dynamical scales and PSs. **Equation 2** denotes a general RSS independence requirement whereby matching is done at a PS $Q^*$ at the $r$-loop, via an effective RSS-independent dynamical coupling and mass depending on a PS, which are denoted as $a_{eff}$ and $m_{eff}$, respectively, in both equations, although in the RSS-dependent expression, these are computed via a truncated order, $n$, thereby denoting the effective nature of these RSS independence conditions. Scale independence is achieved via matching at a physical point $Q^*$, whereby the couplings and masses are referenced to observed values, which ultimately renders the ETOs free from unphysical scale ambiguity. Since POEM is based on the EFT matching approach and focused directly on the observable itself, it must also be noted that in this approach, the PS or reference matching scale $Q^*$ is not arbitrary but is based on the physical observable, the relevant physical process, and the relevant and valid EFT describing the physics process with relevant degrees of freedom. The value of the coupling (and masses) must be experimentally determined at this physical scale, and the range of applicability cannot exceed beyond the relevant physical degrees of freedom or go beyond the perturbative cutoff of the theory. Moreover, the resummation that happens due to POEM, in analogy, "integrates out" the unphysical RSS dependencies from the observable to an independent ETO. In regard to the arbitrary renormalization scale $\mu$ dependence, by matching and referencing the EFT at a physical scale $Q^*$ at which the coupling (and masses) is determined via experiments, POEM leads to scale independence in the observable as the observable is now fully renormalized to the dynamical scale $Q$. As such, with POEM, we also perform an "effective dynamical renormalization" (EDR) of the RSS-dependent physical observable via matching (and resulting resummation) which renders it completely dynamical, with the ETO tied to the physical relevance of the scale of matching with applicability to the underlying theory while preserving physical degrees of freedom.

Scheme independence in $a_{eff}$ and $m_{eff}$ and ultimately the ETO is simultaneously achieved using matching, whereby explicit dependence is absorbed into these running parameters and implicit dependence is eliminated by restricting their running to $r$-loops; hence, we use truncated RG functions at this order. In **Equation 2**, the $k$ subscript denotes a particular scheme for a truncated observable $O$ at a loop order $n$ (for example, the $\overline{MS}$ scheme) which is matched at the $r$-loop level with an effective version of the fully PS-dependent version $O_{eff}$, whereby $r$ is to be chosen at the highest loop order at which scheme independence holds. The rationale behind this requirement is that this allows matching at a perturbative order at which this scheme independence requirement holds fully at the highest allowed order of perturbation theory; hence, results derived are then more accurate than those at lower orders where scheme independence may still hold. Typically, for most observables, $r$ is either at one loop or two loops of perturbation theory. In our study of QCD observable $R_{e^+e^-}$, which is scheme-independent at two loops (in mass-independent renormalization schemes such as $\overline{MS}$), however, we do compute the results at both $r = 1$ and $r = 2$ loop orders and demonstrate as to why matching at the latter order yields different and better results.

Typically, EFT techniques are applied at the level of operators instead of observables and deal with physical degrees of freedom, dealing with RSS dependence as auxiliary variables. Therefore, broadly speaking, our approach via POEM is to bridge conventional renormalization techniques with EFT techniques, with a first focus on observables in this work. Since POEM and the matching of observables to achieve RSS independence are unprecedented, we demonstrate its efficacy in this approach first and we will explicate the construction of ETOs from the RSS-independent Lagrangian of the underlying EFT in a separate work[1]. The advantage of using our approach over using fixed-order perturbation theory is that both explicit dependence and implicit dependence on the renormalization scale and scheme are eliminated at an order-by-order basis via the POEM matching process, and the derived ETOs are resummed expressions that are referenced at a physically relevant matching scale.

Overall, **Eq. 2** allows for a practical realization of **Eq. 1** at a finite order of perturbation theory for our proposed implementation of the POEM approach and implies RSS independence in the ETO to hold both at the matching scale $Q^*$ and for dynamical degrees of freedom, $Q$ and $M$, which can be explicitly stated as follows:

$$\frac{\partial O_{eff}^{(r)}}{\partial \mu} = \frac{\partial O_{eff}^{(r)}}{\partial c_i} = 0, \qquad (3)$$

where $\mu$ is the renormalization scale, while $c_i$ are renormalization scheme-dependent RG coefficients.

We remark here that **Eq. 3** is due to the application of POEM and the resulting resummed expressions, or ETOs, $O_{eff}^{(r)}$. In contrast, the approach of PMS [1–4] is applied to fixed-order perturbative expressions to find optimal scales. Also, our approach is distinct from the effective charge (EC) approach [5, 6] as in this method, there is matching done at one loop and at a scale in which ultraviolet logarithms are set to zero, which only allows renormalization scheme independence, while renormalization scale dependence remains in resulting expressions. Via POEM and for deriving a resultant ETO, the matching at the tree and at the $r$-loop order at a PS, we overcome these limitations, thereby achieving RSS independence simultaneously. In contrast, as remarked above, in PMS, an optimal scale is found separately from the renormalization-scheme invariants. As such, we emphasize that POEM is a distinctly new principle which deals with RSS dependencies simultaneously. Moreover, POEM-based results are RSS-independent; hence, there are no "commensurate scale relations" [26, 27] as the observable is dynamical with respect to physical scales, and hence, no relative scales exist for the ETO. After achieving RSS invariance, there is no issue of renormalons [28, 29] encountered as well. In the case of the QCD cross-section ratio, $R_{e^+e^-}$ detailed in the next section, we have an observable which is scheme-independent up to two loops, and we find that matching at this higher loop yields better results as compared to one-loop matching; hence, we demonstrate the advantages of matching at a higher loop order with RSS-independent results unlike the EC approach, which are scale-dependent and are fixed at the one-loop order for all observables.

For the case of the cross-section ratio $R_{e^+e^-}$, for the $e^+e^- \to$ hadrons studied here, the matching scale $Q^*$ is chosen at the Z-pole mass with five active quarks not only because the strong coupling constant is determined via experiments at this PS in the $\overline{MS}$ scheme but also because at the center of mass energies, $Q$ under consideration is well above the b-quark threshold mass and is within the limits of applicability of the QCD as an EFT for six quark flavors. It must be emphasized that in the case of QCD, the strong coupling constant in $\overline{MS}$ is determined at the Z-pole mass for the reason that at this physical scale, it is most accurate due to the EFT representing free quarks and gluons (rather than at lower energies), and hence, a clean determination of this strong coupling constant is most feasible here; hence, this choice is for QCD, and the motivation for POEM yields physical RSS-independent results. More generally, when a physical reference matching scale $Q^*$ and experimentally determined coupling are chosen and matching is done via POEM, this ends $\mu$ dependence altogether. Hence, with this matching to relevant physical scales, there is no dependence of the ETOs on $Q^*$ itself. This is expressed as follows:

$$\frac{\partial O_{eff}^{(r)}}{\partial Q^*} = 0. \quad (4)$$

---

[1]Farrukh A. Chishtie, in preparation.

It follows that if there is a different matching scale chosen based on the initial physical matching reference scale such that

$$Q^* \to Q^{*\prime} = kQ^*. \quad (5)$$

Then, the dynamical scale must also rescale by the same constant $k$ as follows:

$$Q \to Q' = kQ. \quad (6)$$

This is consistent with **Eq. (4)**, or in other words, this is consistent with the reference physical scale of $Q*$ (and the underlying measurement of couplings and masses at this scale). While renormalization group (RG) functions can be used to evolve the coupling and masses using $Q*$ to another matching scale, $Q^{*\prime}$, however, the ETO results would be invariant to this shift as the reference matching scale has not changed (as it used as an initial value of the coupling evolution), and any rescaling of this physical scale would then result in a compensatory rescaling in $Q$ as indicated by **Eq. 6**. The RSS-independent ETO derived will be, of course, limited in terms of its range of applicability given the reasons for the choice of $Q*$, and here, we explore this in detail with QCD as an underlying EFT with physical energies above the b-quark mass. As another example, if a physical observable related to lower energies is taken much below this threshold, then the matching scale $Q*$ can be chosen at which the strong coupling is experimentally measured at the tau mass and for three to four active quark flavors.

Finally, it is interesting to note that since RSS independence is guaranteed trivially at the tree level so that via **Eq. 2**, we also can express this particular case as follows:

$$O_n^{(k)}(a_{eff}(\Lambda_{eff}), m_{eff}(\Lambda_{eff}), \Lambda_{eff}, Q, M) = O_{eff}^{(tree)}. \quad (7)$$

In this relationship, we introduce a dynamical scheme-independent cutoff scale at renormalization $\mu = \Lambda_{eff}$ for the observable which denotes the point at which no quantum corrections occur in the matching RSS-independent theory at the tree level. As it is dependent only on dynamical scales, this RSS-independent cutoff scale is indeed distinct from typical renormalization coupling and mass cutoffs, which are scheme-dependent (see [25] for analysis on RS dependence of the strong coupling constant and its cutoff $\Lambda_{QCD}$). As such, **Eq. 6** is a particular case of the general POEM condition, which has interesting new implications, while holding no bearing in deriving ETOs which necessarily contain perturbative quantum corrections.

## 3 ATTAINING EFFECTIVE RENORMALIZATION SCALE AND SCHEME INDEPENDENCE VIA MATCHING AT THE ONE-LOOP ORDER

The relation of the cross-section ratio $R_{e^+e^-}$ is given by $3(\sum_i q_i^2)(1+R)$, where $R$ at the $n$-loop order has a perturbative contribution of order $a^{n+1}$ in

$$R = R_{\text{pert}} = \sum_{n=0}^{\infty} r_n a^{n+1} = \sum_{n=0}^{\infty} \sum_{m=0}^{n} T_{n,m} L^m a^{n+1}, \quad (8)$$

with $L \equiv b \ln(\frac{\mu}{Q})$, $Q^2$ being the center of mass energy squared.

The explicit dependence of $R$ on the renormalization scale parameter $\mu$ is compensated for by implicit dependence of the "running coupling" $a(\mu^2)$ on $\mu$,

$$\mu^2 \frac{\partial a}{\partial \mu^2} = \beta(a) = -ba^2(1 + ca + c_2 a^2 + \ldots), \quad (9)$$

where $a \equiv \alpha_s(\mu^2)/\pi$, while $\alpha_s$ is the QCD strong coupling constant.

The cross-section ratio $R_{e^+e^-}$ for five active flavors of quark is as follows:

$$\frac{3}{11} R_{e^+e^-} = 1 + R_{\text{pert}}, \quad (10)$$

where the choice of the number of flavors is based on the center of mass energies $Q$ (for which the ETO will be derived) which are for those above the b-quark mass.

In the $\overline{MS}$ renormalization scheme [30], we have

$$b = 23/12, \quad c = 29/23, \quad c_2 = 9769/6624, \quad c_3 = 9.835917120, \quad (11)$$

where the values of $b$ and $c$ are the same in any mass-independent renormalization scheme, while the values of $c_2$ and $c_3$ in **Eq. 11** are particular to the $\overline{MS}$ scheme. Furthermore, we find in Refs. [31, 32] that in the $\overline{MS}$ scheme,

$$\begin{aligned} T_{0,0} &= 1, \ T_{1,0} = 1.4097, \ T_{1,1} = 2 \\ T_{2,0} &= -12.76709, \ T_{2,1} = 8.160539, \ T_{2,2} = 4 \\ T_{3,0} &= -80.0075, \ T_{3,1} = -66.54317, \ T_{3,2} = 29.525095 \ T_{3,3} = 8. \end{aligned} \quad (12)$$

At the Z-pole mass ($M_Z = 91.1876\ GeV$), we have [33] in the $\overline{MS}$ scheme (with $a(\mu)$ governed by **Eq. 9**),

$$\pi\, a_{\overline{MS}}(M_Z) = 0.1179 \pm 0.001. \quad (13)$$

Utilizing **Eq. 2**, we therefore find the following POEM-based relationship:

$$\begin{aligned} \frac{3}{11} R_{e^+e^-}^{\overline{MS}}(Q^*, Q) &= 1 + R_{\text{pert}}(Q^*, Q) = \frac{3}{11} R_{e^+e^-}^{eff}(Q^*, Q) \\ &= 1 + a_{eff}^{1L}(Q^*, Q). \end{aligned} \quad (14)$$

With **Equation 14**, which is matching done at the one-loop level, we choose the matching point $Q* = M_Z = 91.1876 GeV$ and also subtract the only explicit scheme-dependent term, appearing in the expression for $T_{31}$, which is $2c_2$ in the $\overline{MS}$ expression, rendering the overall matching expression RSS-independent. With the central value of the strong coupling $a^{\overline{MS}} = \frac{0.1179}{\pi}$ from **Eq. 13**, which we substitute into **Eq. 14**, we derive $a_{eff}^{1L}(M_Z, Q)$ as follows:

$$a_{eff}^{1L}(M_Z, Q) = 0.03868 + 0.0059614 \ln\left(\frac{M_Z}{Q}\right)$$
$$+ 0.0009918 \ln^2\left(\frac{M_Z}{Q}\right) + 0.0001117 \ln^3\left(\frac{M_Z}{Q}\right). \quad (15)$$

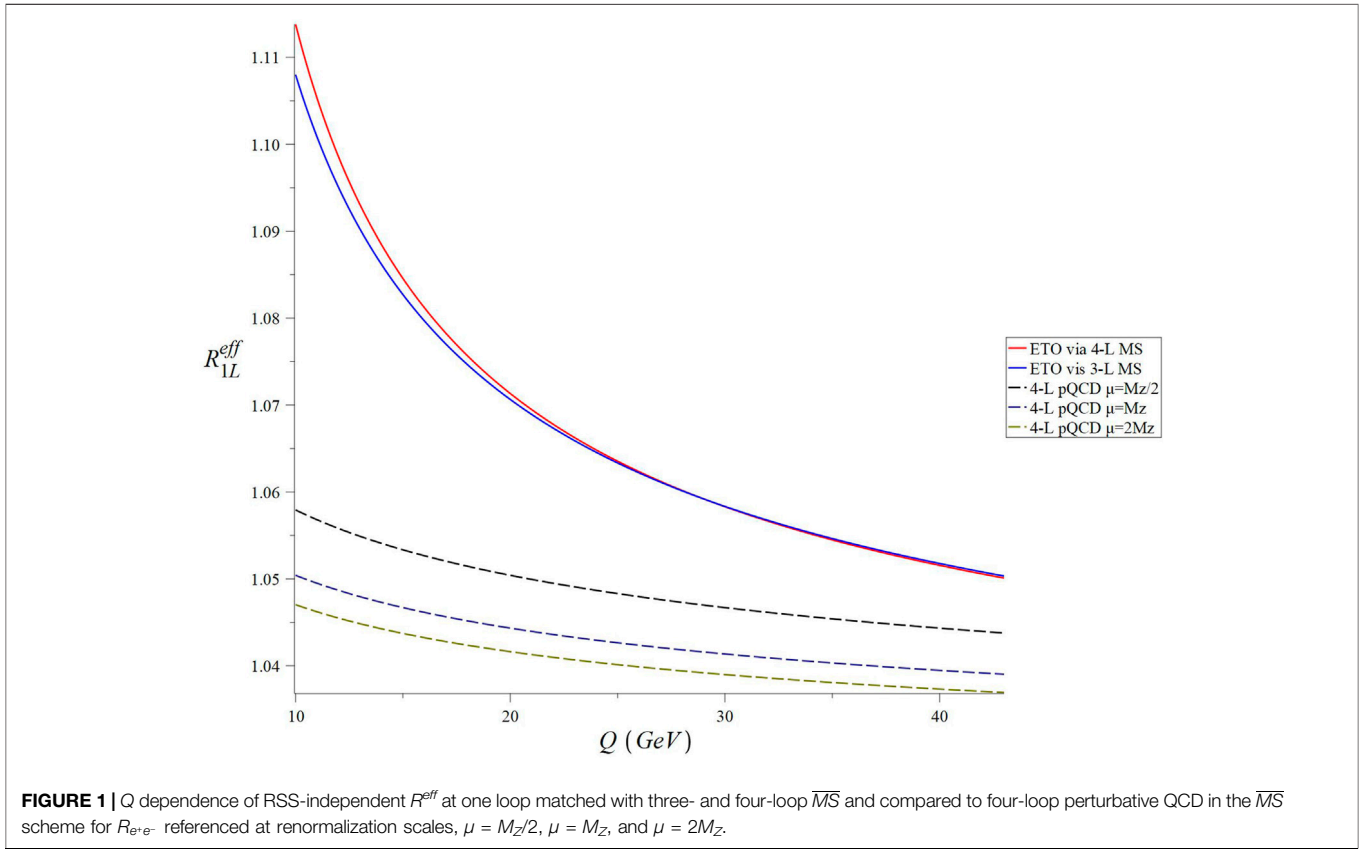

**FIGURE 1** | $Q$ dependence of RSS-independent $R^{eff}$ at one loop matched with three- and four-loop $\overline{MS}$ and compared to four-loop perturbative QCD in the $\overline{MS}$ scheme for $R_{e^+e^-}$ referenced at renormalization scales, $\mu = M_Z/2$, $\mu = M_Z$, and $\mu = 2M_Z$.

The choice of $Q* = M_Z$ and the number of quarks, $n_f = 5$, is predicted on deriving the ETO for center of mass energies $Q$ above the b-quark mass (and also below the t-quark threshold mass).

Since for the ETO we absorb higher-order loop contributions from an RSS-dependent scheme at the one-loop order for RSS independence, $R_{e^+e^-}^{eff}(Q)$ is expressed as

$$\frac{3}{11}R_{e^+e^-}^{eff}(Q) = 1 + a_{eff}^{1L}(Q), \quad (16)$$

where

$$a_{eff}^{1L}(Q) = \frac{a_{eff}^{1L}(Q*, Q)}{1 - ba_{eff}^{1L}(Q*, Q)\ln\left(\frac{Q*^2}{Q^2}\right)} \quad (17)$$

is a one-loop beta-function solution in the dynamical ETO, with an initial value set at $a_{eff}^{1L}(\mu_0 = Q*) = a_{eff}^{1L}(Q*, Q)$. This, in our case, is given by **Eq. 15**.

Since we are at the one-loop matching, we find the one-loop solution to $a_{eff}(Q)$ which leads to the prediction of $\frac{3}{11}R_{e^+e^-}^{eff} = 1.056943^{+0.0007}_{-0.0007}$ from **Eq. 16** at $Q = 31.6 GeV$, which is in excellent agreement with the experimental value of $\frac{3}{11}R_{e^+e^-}^{exp} = 1.0527^{+0.005}_{-0.005}$ [34]. We plot the $Q$ dependence of $R^{eff}$ at one loop matched with three- and four-loop $\overline{MS}$ in **Figure 1**, where both results are similar. At the same time, perturbative QCD (pQCD) predictions are found for renormalization scales, $\mu = M_Z/2$, $\mu = M_Z$, and $\mu = 2M_Z$, respectively, which are all underestimates as compared to POEM-based ETO results, and also indicate high uncertainty across this range due to RSS dependence and resulting ambiguities.

## 4 TWO-LOOP MATCHING

$R_{e^+e^-}$ is scheme-independent at two loops; hence, this would be applying POEM at the maximum loop order in this case. Utilizing **Eq. 2** at the two-loop level, we find the following scale- and scheme-independent POEM-based relation:

$$\frac{3}{11}R_{e^+e^-}^{\overline{MS}}(Q*, Q) = 1 + R_{pert}(Q*, Q) = \frac{3}{11}R_{e^+e^-}^{eff}(Q*, Q)$$
$$= 1 + a_{eff}^{2L}(Q*, Q) + (T_{1,0} + T_{1,1}L)(a_{eff}^{2L}(Q*, Q))^2. \quad (18)$$

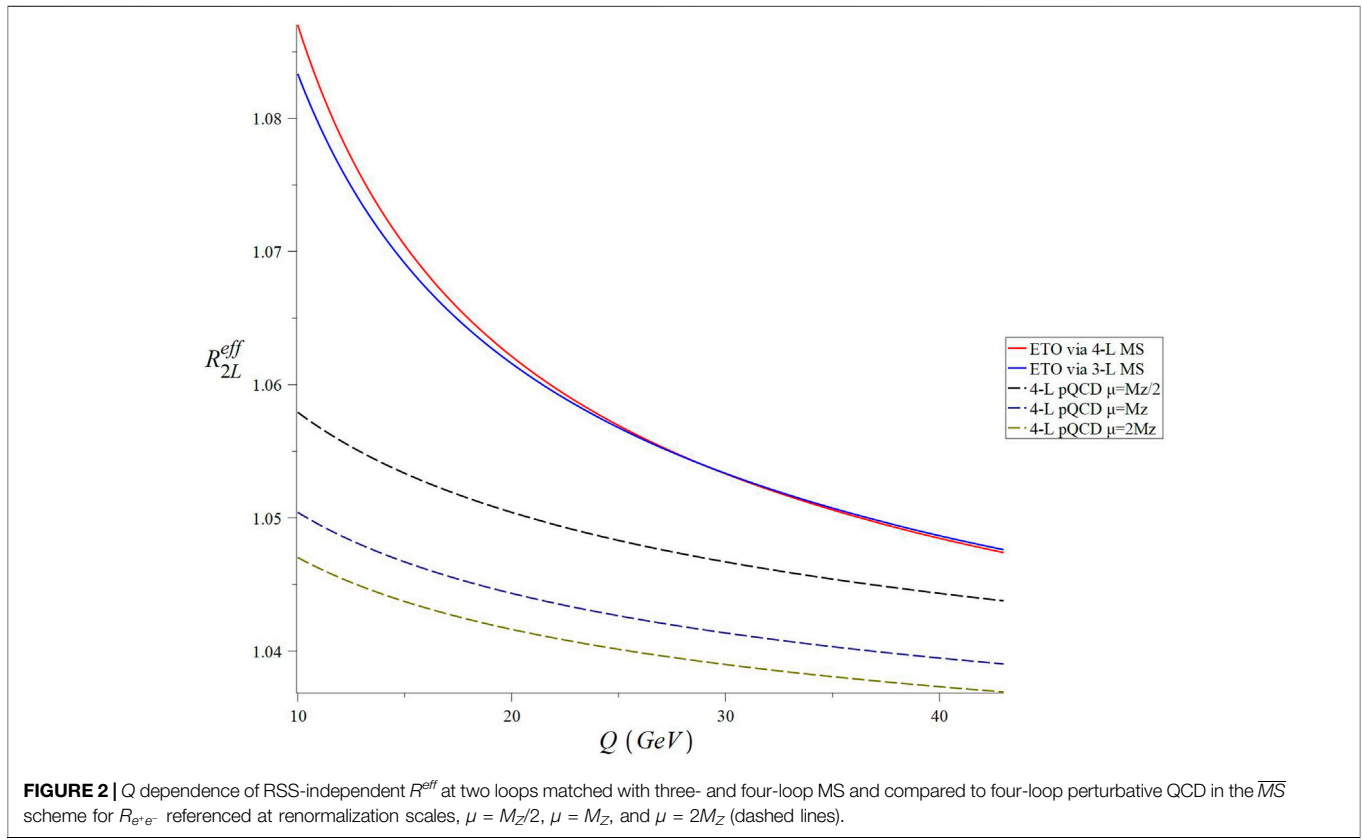

**FIGURE 2** | $Q$ dependence of RSS-independent $R^{eff}$ at two loops matched with three- and four-loop MS and compared to four-loop perturbative QCD in the $\overline{MS}$ scheme for $R_{e^+e^-}$ referenced at renormalization scales, $\mu = M_Z/2$, $\mu = M_Z$, and $\mu = 2M_Z$ (dashed lines).

As in the previous section, we choose $Q* = M_Z$, eliminate the scheme-dependent term, and find $a_{eff}^{2L}(M_Z, Q)$ from solving **Eq. 18** and obtain an RSS-independent matching expression:

$$a_{eff}^{2L}(M_Z, Q) = \pm k_1 \left[ \mp 1 + \left\{ 1.2181 + 0.0008939 \ln^4\left(\frac{M_Z}{Q}\right) \right. \right.$$
$$\left. \left. + 0.008565 \ln^3\left(\frac{M_Z}{Q}\right) + 0.05328 \ln^2\left(\frac{M_Z}{Q}\right) + 0.3431 \ln\left(\frac{M_Z}{Q}\right) \right\}^{1/2} \right]. \quad (19)$$

Here, $k_1 = [2\{1.4097 + 2\ln(\frac{M_Z}{Q})\}]^{-1}$. Since in the ETO, namely, $R_{e^+e^-}^{eff}$, we absorb higher-order loop contributions from an RSS-dependent scheme at $r = 2$ (two-loop orders) for RSS independence, $R_{e^+e^-}^{eff}(Q)$ is expressed as follows:

$$\frac{3}{11} R_{e^+e^-}^{eff}(Q) = 1 + a_{eff}^{2L}(Q) + T_{1,0} a_{eff}^{2L}(Q)^2, \quad (20)$$

where

$$a_{eff}^{2L}(Q) = -\frac{1}{c\left[W_{-1}\left(-\frac{\exp(f)}{c}\right) + 1\right]} \quad (21)$$

and

$$f = \frac{a_{eff}^{2L}(Q, Q^*)\left[b \ln\left(\frac{Q^{*2}}{Q^2}\right) + c \ln(h) - c\right] - 1}{c a_{eff}^{2L}(Q, Q^*)}, \quad (22)$$

while

$$h = \frac{c a_{eff}^{2L}(Q, Q^*) - 1}{a_{eff}^{2L}(Q, Q^*)}. \quad (23)$$

**Equations 21–23** represent an exact closed form solution of the two-loop beta-function for the ETO, which is expressed as a Lambert-$W$ function, whereby $W_{-1}(\zeta)$ denotes the applicable branch of the function for relevant values of $\zeta > 1$. We utilize the initial value set at $a_{eff}(\mu_0 = Q*) = a_{eff}(Q*, Q)$, which in our case is given by the positive root of **Eq. 20**. At the two-loop ETO matching, we find that the prediction is nearly the same as the one-loop ETO, which is $\frac{3}{11} R_{e^+e^-}^{eff} = 1.052431_{-0.0006}^{+0.0006}$ from **Eq. 21** at $Q = 31.6 GeV$, which is in excellent agreement with the experimental value of $\frac{3}{11} R_{e^+e^-}^{exp} = 1.0527_{-0.005}^{+0.005}$ [34]. This is better than our previously derived one-loop matching ETO result.

We plot the $Q$ dependence of the ETO $R^{eff}$ at two-loop matching with three- and four-loop $\overline{MS}$ in **Figure 2** and find the same trends as one-loop matching, with pQCD predictions lower than ETO results, although these are slightly lower than the one-loop ETO results. We also show the behavior of RSS-independent coupling $a^{eff}$ at one- and two-loop orders with respect to $Q$, in **Figure 3**, which depict asymptotic freedom as is also seen in the ETO behavior in **Figures 1**, **2**, with both three- and four-loop $\overline{MS}$ results close to each other, both at one- and two-loop ETO matched results, and within the experimental error bounds for the $Q = 31.6 GeV$ experimental value. We note that the effective dynamical coupling $a^{eff}$ is lower for two loops as compared to one-loop matching as depicted in **Figure 3**, which results in a better prediction of $\frac{3}{11}R^{eff}_{e^+e^-}$. In **Figure 4**, we show that at two loops $r = 2$, which is the highest loop order at which $\frac{3}{11}R_{e^+e^-}$ is scheme-independent, these ETO predictions are lower (and subsequently better) than those found at one loop. Both results derived here are within the experimental errors with the two loops ($r = 2$), providing much better agreement with data as compared to the one-loop ETO. Based on these findings, we therefore surmise matching at the highest available order at which the observable is scheme-independent to yield the best results.

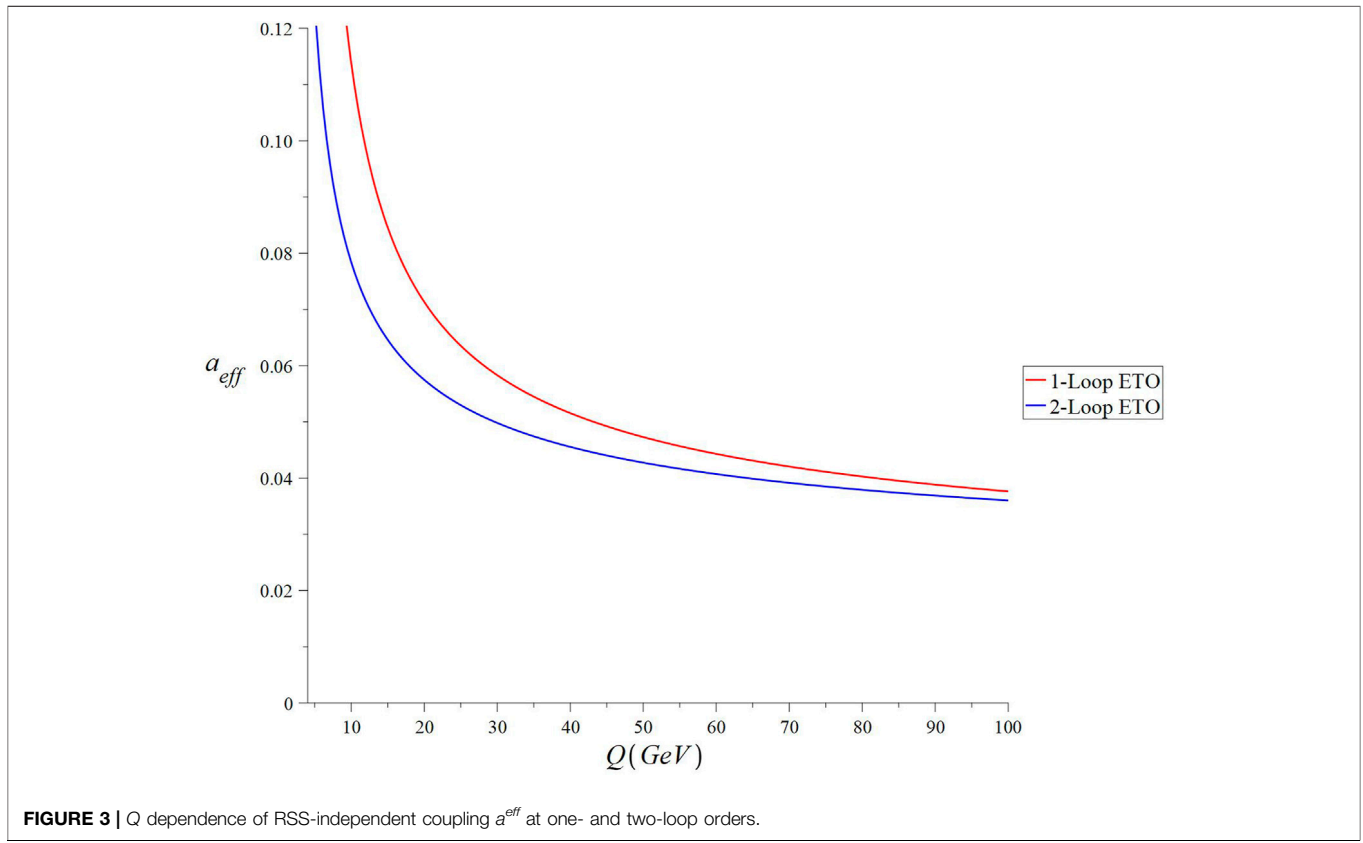

**FIGURE 3** | $Q$ dependence of RSS-independent coupling $a^{eff}$ at one- and two-loop orders.

With respect to considerations of higher-order perturbative corrections, we find that both the one- and two-loop ETOs yield results which are nearly identical for both three- and four-loop $\overline{MS}$ expressions. This is shown in both **Figures 1**, **2**. For $Q = 31.6 GeV$, for the central value of the strong coupling constant, for the two-loop ETO, we find that for three loops, $\frac{3}{11}R^{eff}_{e^+e^-} = 1.057040$, which is 0.006% higher than the four-loop prediction of 1.052431 (stated earlier). Using comparison of three-loop results with four-loop results for both one- and two-loop ETOs indicates that the ETOs derived via POEM are highly convergent (since these are free from RSS dependence and also from renormalons). Hence, higher-order corrections in $\overline{MS}$ as in five-loop orders and beyond would contribute negligibly for the highly convergent ETOs derived in this work.

In comparison with our findings, Akrami and Mirjalili [37] present perturbative QCD, RG summation and RS invariants, and CORGI approach estimates of $R$ at $Q = 31.6 GeV$ to be $1.04617^{+0.0006}_{-0.0006}$, $1.04711^{+0.00003}_{-0.00005}$, and $1.04615^{+0.0015}_{-0.0008}$ at four loops, respectively, which are all underestimates and fall outside of the experimental error bounds, in contrast to the results derived by POEM. This pattern holds for other experimental measured values at $Q = 42.5 GeV$ and $Q = 52.5 GeV$, which are $1.0554 \pm 0.2$ [35] and $1.0745 \pm 0.11$ [36], respectively. With POEM, we find $1.047561^{+0.0005}_{-0.0005}$ and $1.044679^{+0.0005}_{-0.0005}$ using the two-loop ETO matching. These predictions are higher in value and more accurate than those reported for perturbative QCD, RG summation and RS invariants, and CORGI approaches in [37]. However, our predictions are lower than the experimental values as they are based only on photon interactions, and a higher accuracy is expected when electroweak contributions arising from electron–positron annihilation to $Z$-bosons are taken into account. We will address this in a future work, which will also additionally address other processes [39].

## 5 CONCLUSION

In this work, we have introduced a new approach for achieving RSS independence via a principle termed as POEM, which is based on equivalence of physical observables across both physical scale and scheme dependencies in their perturbative content under certain limits. Inspired by EFT techniques, which involve matching at a physical scale, this integrated renormalization approach, via EDR, provides results as RSS-independent ETOs, which contain only physical scales at a fixed order of perturbation theory. We demonstrate that POEM provides excellent results for the QCD cross-section ratio $R^{eff}_{e^+e^-}$, and agreement with experiments is better than other comparable estimates based on fixed-order perturbative QCD, RG summation and RS invariants, and CORGI approaches [37]. Moreover, both the one- and two-loop ETOs show remarkable convergence as both three and four loops provide results which are nearly identical, which indicates the high convergence rate of the ETOs due to the lack of RSS dependence and renormalons. POEM is distinct from present approaches in its conceptualization, but we do intend to find potential connections and improvements with other approaches including $PMC_\infty$ [38] and the recently devised distribution-based approach [39]. In the case study, we have only focused on observables incorporating photon interactions and hence for better accuracy for higher values of $Q$ are expected to have contributions from electron–positron annihilation to the $Z$-boson to be taken into account[1]. Overall, we have generally focused on the observables derived from this full RSS-independent theory via POEM. We further aim to study and explicate POEM's potential in these areas in upcoming studies, whereby linkages between conventional renormalization and EFT techniques are further explicated with a means to find the underlying RSS-independent EFT[2], and as a follow-up will also \address other electroweak processes using POEM including Standard Model Higgs decays and cross-sections[3]. Based on a new conceptualization and achieving better results for the cross-section ratio $R^{eff}_{e^+e^-}$, we contend that POEM is potentially applicable widely to achieve RSS independence in physical observables across the high-energy physics domain including the standard model (SM) and beyond the SM physics, along with areas such as cosmology and statistical and condensed matter physics, where RSS ambiguities are regularly encountered.

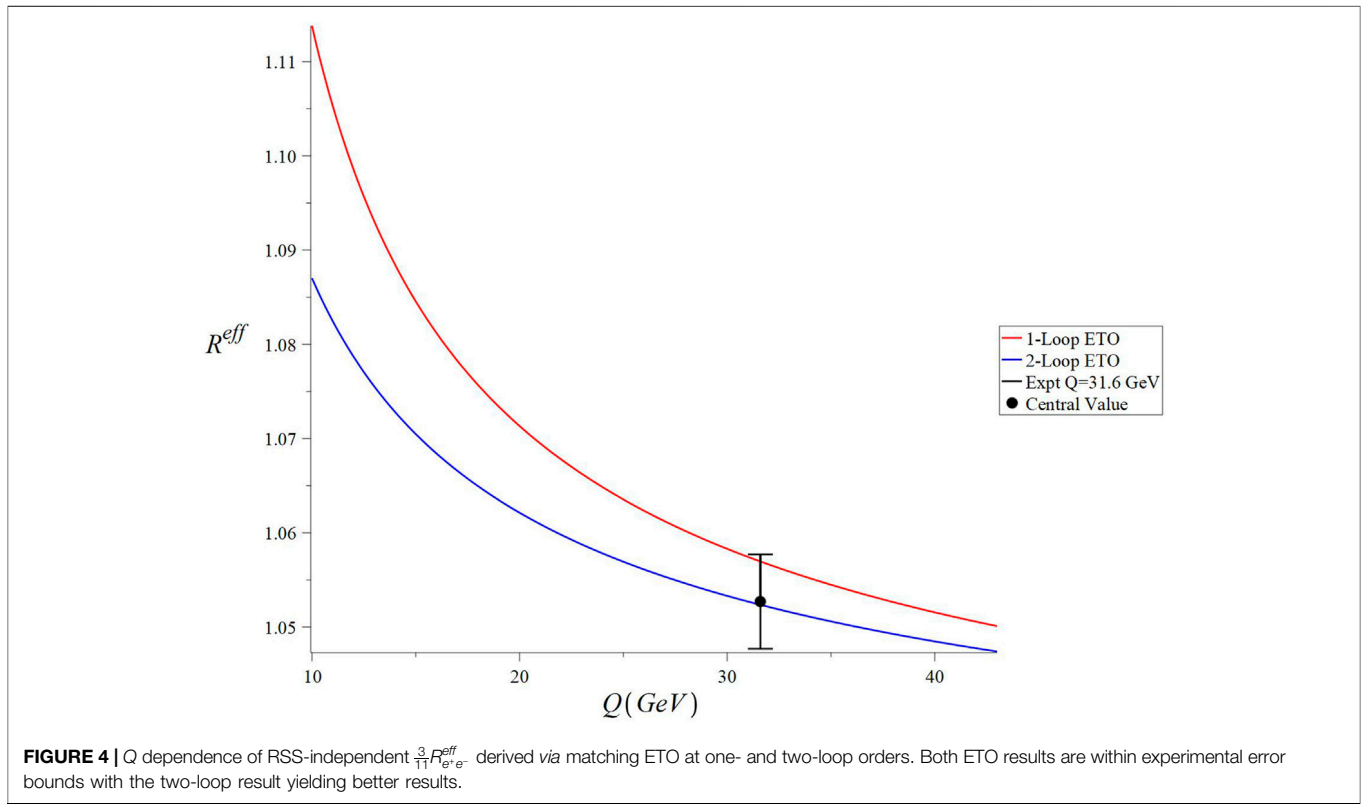

**FIGURE 4 |** $Q$ dependence of RSS-independent $\frac{3}{11}R^{eff}_{e^+e^-}$ derived via matching ETO at one- and two-loop orders. Both ETO results are within experimental error bounds with the two-loop result yielding better results.

## DATA AVAILABILITY STATEMENT

The raw data supporting the conclusion of this article will be made available by the authors, without undue reservation.

---

[2]Farrukh A. Chishtie, in preparation.
[3]Farrukh A. Chishtie, in preparation.

## AUTHOR CONTRIBUTIONS

The author confirms being the sole contributor of this work and has approved it for publication.

## ACKNOWLEDGMENTS

The author thanks Gerry McKeon and Thomas G. Steele for helpful discussions and comments on the initial drafts of this work.